\documentclass{osa-article}

\journal{oe}
 
\articletype{Research Article}

\usepackage{lineno}
\begin{document}

\title{Optimizing the Rydberg EIT spectrum in a thermal vapor}

\author{Hsuan-Jui Su\authormark{1}, Jia-You Liou\authormark{1}, I-Chun Lin\authormark{1}, and Yi-Hsin  Chen\authormark{1,2,*}}
\address{\authormark{1}Department of Physics, National Sun Yat-Sen University, Kaohsiung, 80424, Taiwan\\
\authormark{2}Center for Quantum Technology, Hsinchu 30013, Taiwan}
\email{\authormark{*}Corresponding author: yihsin.chen@mail.nsysu.edu.tw} 



\begin{abstract}
We present Rydberg-state electromagnetically-induced-transparency (EIT) measurements examining the effects of laser polarization, magnetic fields, laser intensities, and the optical density of the thermal $^{87}$Rb medium. 
Two counter-propagating laser beams with wavelengths of 480 nm and 780 nm were employed to sweep the spectrum across the Rydberg states $|33D_{3/2}\rangle$ and $|33D_{5/2}\rangle$. An analytic transmission expression well fits the Rydberg-EIT spectra with multiple transitions under different magnetic fields and laser polarization after accounting for the relevant Clebsch-Gordan coefficients, Zeeman splittings, and Doppler shifts. In addition, the high-contrast Rydberg EIT can be optimized with the probe laser intensity and optical density. Rydberg EIT peak height was achieved at $13\%$, which is more than twice as high as the maximum peak height at room temperature. A quantitative theoretical model is employed to represent the spectra properties and to predict well the optimization conditions.
A Rydberg EIT spectrum with high contrast in real-time can be served as a quantum sensor to detect the electromagnetic field within an environment.
\end{abstract}

\newcommand{\FigOne}{
\begin{figure}[t]
\centering
\includegraphics[width=0.95 \linewidth]{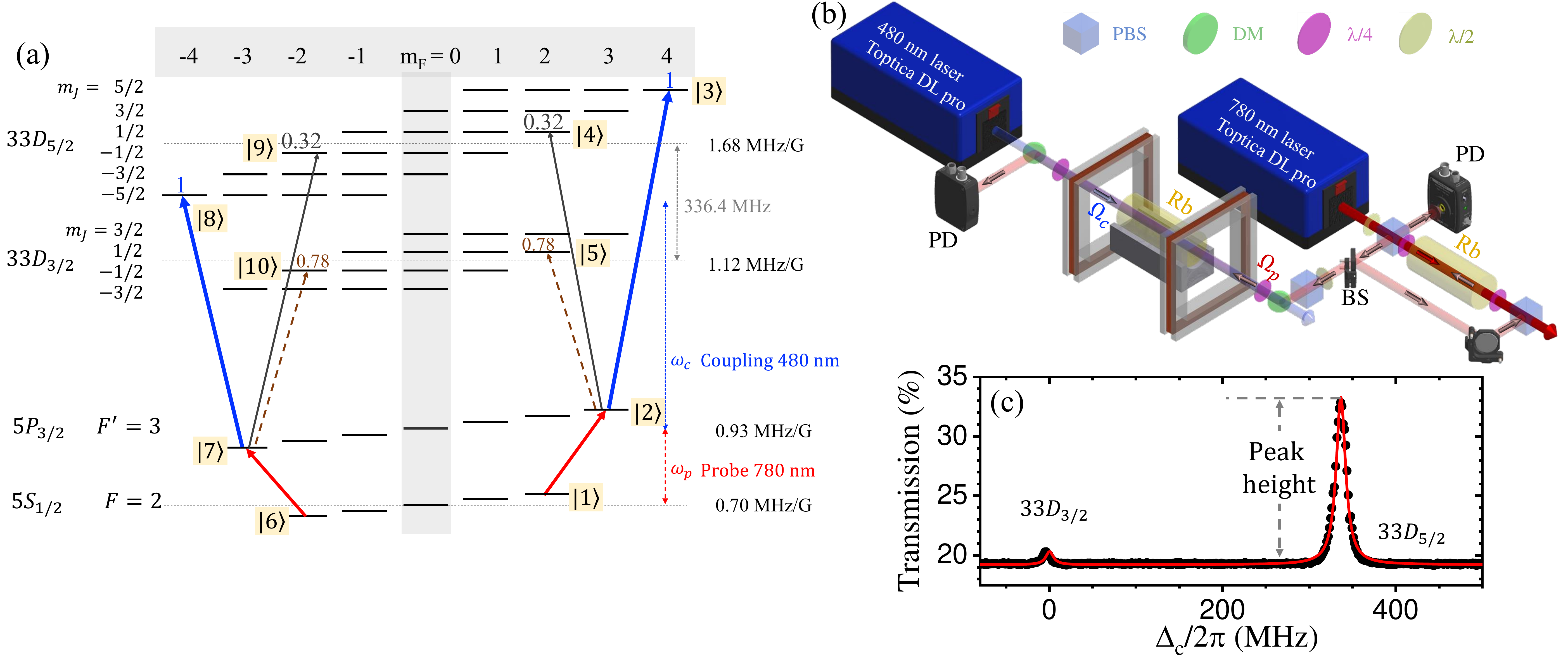}
\caption{(a) Zeeman energy sublevel diagram for $^{87}$Rb atom under a specific magnetic field. The 780-nm probe field (Toptica DL pro) was frequency locked via a reference cell (homemade Rb cell) to the transition of the ground state $|5S_{1/2}, F=2\rangle$ to the intermediate state $|5P_{3/2}, F'=3\rangle$. The 480-nm coupling field (Toptica DL pro HP 480) was swept across the transitions of $|5P_{3/2}, F'=3\rangle$ to $|33D_{3/2}\rangle$ and $|33D_{5/2}\rangle$. The probe $\Omega_p$ and coupling $\Omega_c$ fields form a $\Xi$-type EIT configuration. Due to the optical-pumping effect, we assume the atomic population accumulates in state $|1\rangle$ with $\sigma^{+}$-polarized $\Omega_p$ or equally distributes in states $|1\rangle$ and $|6\rangle$ with linear-polarized $\Omega_p$. The numbers indicate the Clebsch-Gordan coefficients (CGCs) for the transitions between different Hyperfine and Zeeman states. Zeeman splittings among magnetic sublevels are also presented. (b) Schematic of the experimental setup. The two beams were sent counter-propagating to the center of the vapor cell. Two dichroic mirrors (DMs) were applied to combine and separate these two color beams. 
The magnetic field was generated with a pair of rectangular Helmholtz coils. We controlled lasers' polarization by two quarter-wave plates ($\lambda/4$). The full width at $e^{-2}$ maximum of the probe and coupling beams were both around 0.81 mm. The spectra were directly detected by a photodetector (Thorlabs PDA36A2). (c) shows a typical Rydberg-state EIT spectrum taken under the optimum probe intensity of 0.04 W/cm$^2$, $\sigma^+$-$\sigma^+$ laser polarization configuration, and the magnetic field close to zero. The solid curve is the theoretical prediction from Eq.~(\ref{eq:spectrum}) under a parameter set of $\alpha=1.65$, $\{\Omega_c, \gamma\}=2\pi\times\{7.8,5.4\}$~MHz.}
\label{fig:scheme}
\end{figure}
}

\newcommand{\FigTwo}{
\begin{figure}[t]
\centering
\includegraphics[width=0.95 \linewidth]{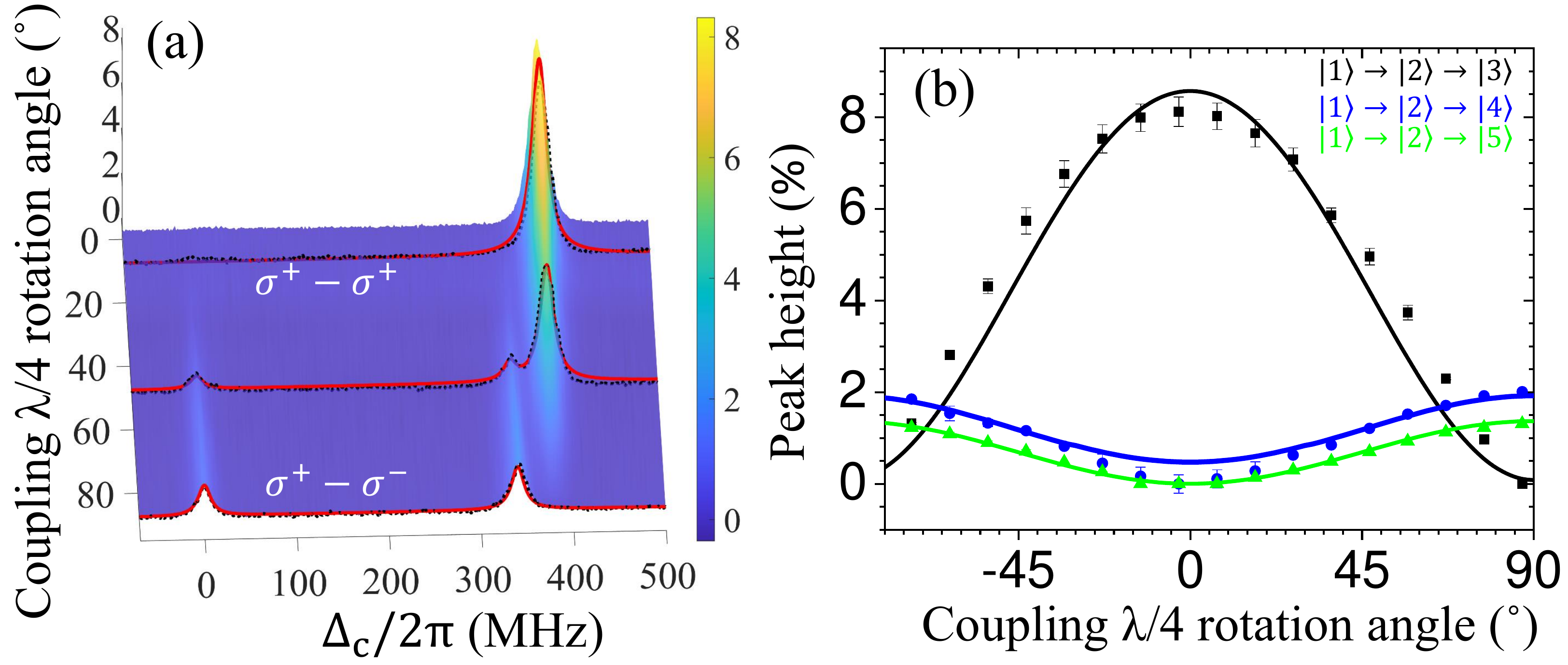}
\caption{(a) Rydberg-state EIT evolution along with the coupling field polarization under a magnetic field of 10.9 Gauss and temperature of $51^{\circ}$C. In the whole measurements, the probe field was maintained in the $\sigma^+$-polarization. The spectra with $\sigma^+$, linear, and $\sigma^-$ coupling field polarization are also shown in the panel, from the top to bottom. We extract the heights of the three EIT peaks and plot them in (b). 
Black squares, blue circles, and green triangles show the EIT peak heights involving the transitions of $|1\rangle\rightarrow|2\rangle\rightarrow|3\rangle$,  $|1\rangle\rightarrow|2\rangle\rightarrow|4\rangle$, and $|1\rangle\rightarrow|2\rangle\rightarrow|5\rangle$, respectively. 
The solid lines are the simulation results under a set of parameters $\alpha=1.55$, $\{\Omega_c$ (for CGC=1), $\gamma\}=2\pi\times\{5.2,8\}$ MHz, and the calibrated coupling Rabi frequencies for $|33D_{3/2}\rangle$ EIT and left peak EIT of $|33D_{5/2}\rangle$. Details can be found in the main text. Because of the optical pumping by the $\sigma^+$ probe field, we assume all populations are accumulated at state $|1\rangle$. In the simulation, we consider all possible transitions involving states $|1\rangle$ and $|2\rangle$. The relevant CGCs, the Zeeman splitting, the Doppler shift, and the polarization ratio of $\sigma^+$ to $\sigma^-$ of the coupling field are taken into account.}
\label{fig:2}
\end{figure}
}

\newcommand{\FigThree}{
\begin{figure}[t]
\centering
\includegraphics[width=0.95 \linewidth]{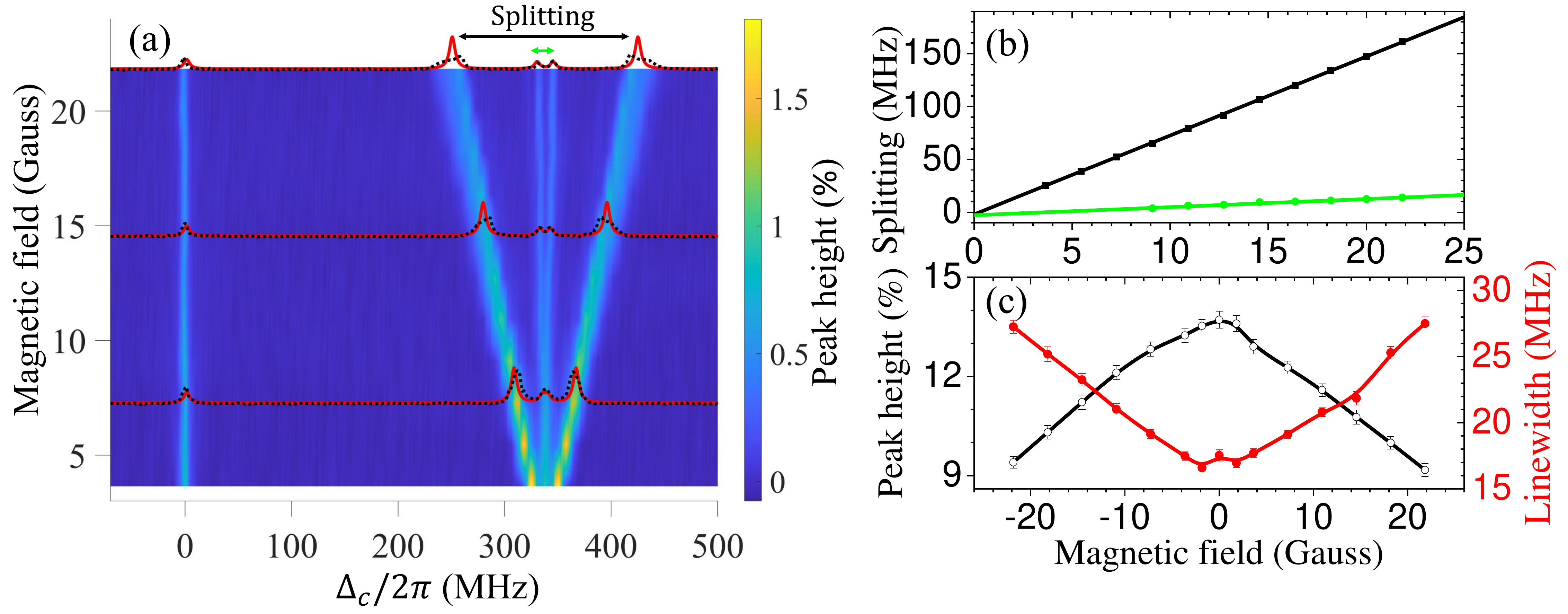}
\caption{(a) Rydberg-state EIT evolution along with the magnetic field under a temperature of $27^{\circ}$C. Both the probe and coupling fields are linearly polarized, which is perpendicular to the quantum axis defined by the magnetic field direction. As the applied magnetic field is sufficiently large, e.g. 6 Gauss, the outermost peaks can be completely resolved. In the panel, we also show the spectral lines for 7.3, 15, and 22 Gauss magnetic fields. The right and left main peaks involving $|33D_{5/2}\rangle$ state corresponds to the transition of $|1\rangle \rightarrow |2\rangle \rightarrow |3\rangle$ and $|6\rangle \rightarrow |7\rangle \rightarrow |8\rangle$, respectively. We take the Zeeman effect and Doppler shift into account in the simulation, shown as the solid lines. The set parameters are $\alpha=0.36\pm 0.02$ according to the baseline transmissions and $\{\Omega_c$ (for CGC=1), $\gamma\}=2\pi\times\{2.1, 6\}$MHz based on the EIT profiles. (b) The frequency splittings of two main EIT peaks (black squares) and two inner peaks (green circles) versus the magnetic field. The linear fits give the frequency splittings of 7.5 and 0.77 MHz/G. (c) The EIT peak height and its relevant EIT linewidth under $\sigma^+$-$\sigma^+$ polarization configuration and temperature of $51^{\circ}$C. The maximum peak height, as well as the narrowest linewidth, occurred when the magnetic field was around zero.}
\label{fig:3}
\end{figure}
}
\newcommand{\FigFour}{
\begin{figure}[t]
\centering
\includegraphics[width=0.95 \linewidth]{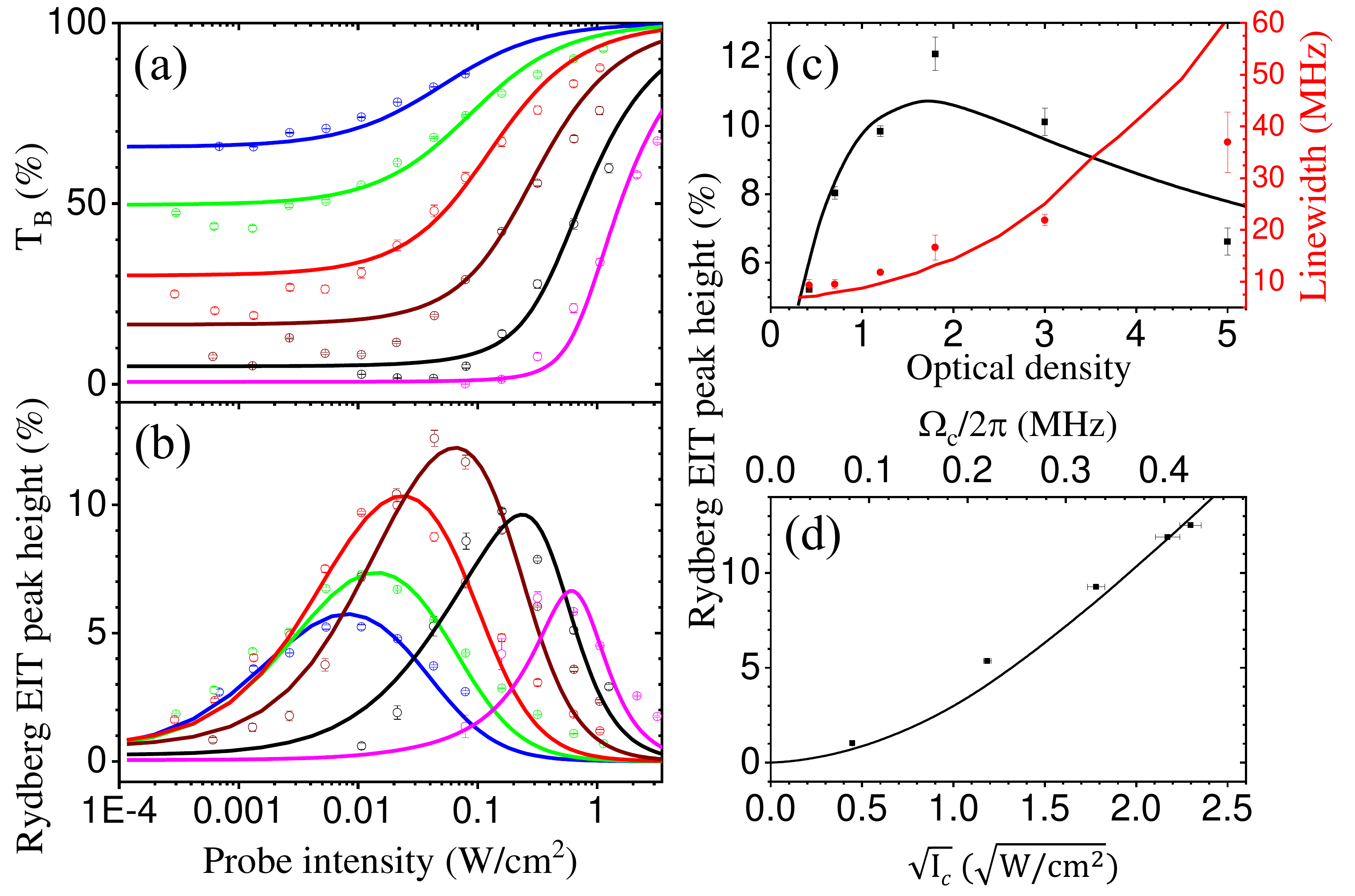}
\caption{(a) Baseline transmission versus the probe field intensity, $I_p$, for vapor temperatures ranging from $27^\circ$C to $65^\circ$C (from top to bottom circles). The transmission raises with $I_p$ mainly due to the power broadening effect. The solid lines are the fits to extract OD and the conversion ratio of laser intensity to $\Omega_p/\Gamma_e$. (b) Peak height of Rydberg EIT versus $I_p$ as a function of vapor temperature (corresponding to the same color in (a)). After the optimization of the laser intensity and OD (varied by the vapor temperature), the maximum Rydberg EIT peak height of $13\%$ was reached when the set vapor temperature was $51^\circ$C. As the OD increases, the optimum $I_p$ becomes stronger, leading to a wider EIT linewidth. The maximum peak height and its relevant linewidth at different ODs are shown in (c). (d) The peak height was monotonously increased with the coupling field intensity, $I_c$. The data were taken under the best experimental conditions, i.e., T$=51^\circ$C and $I_p=0.044$ W/cm$^2$. Therefore, the peak height is not limited to the present result but can be further enhanced by increasing $I_c$. In the whole simulation, $\Gamma_r/2\pi = 5$ kHz and $\gamma/2\pi = 0.75$ MHz. The set $\Omega_c/2\pi$ is $0.38\pm 0.03 $ MHz in (b) and that is fixed as 0.38 MHz in (c). }
\label{fig:ODeffect}
\end{figure}
}

\section{Introduction}
Rydberg atoms have large transition dipole moments between Rydberg states, making them sensitive to the electromagnetic field, among the range of radio-frequency~\cite{Fan2015,Jiao2016,Simons2018,Robinson2021}, microwave~\cite{Sedlacek2012,Fan2014,Jia2020,Meyer2021}, and terahertz~\cite{Wade2017,Wade2018,THz2020}. Rydberg transition allows efficient THz-to-optical conversion, creating high-speed THz videos~\cite{Wade2017,Wade2018,THz2020}.
The rich Rydberg energy levels and external field electrometry can be directly nondissipative detected via electromagnetically-induced-transparency (EIT) spectral measurement~\cite{Mohapatra2007,Shanxia2016,Wu2017,Zhang2018,Xue2019}. The EIT, a quantum interference effect, results in a narrow resonance feature with high accuracy and precision so that it can diagnose energy level shifts caused by the external fields or by the internal atomic interaction. The Rydberg EIT spectrum has been applied to study the Rydberg-atom interaction as well as the cooperative optical nonlinearity in dense cold~\cite{Pritchard2010,coldREIT2013} or thermal ensembles~\cite{Mohapatra2008,Baluktsian2013}. Consequently, the Rydberg-atom-based calibration-free field measurements have potential applications in sensing and communication~\cite{Song2019,Fancher2021}. 

A high-contrast Rydberg-EIT spectrum with a portable non-metal vapor cell will advance the above-mentioned applications. 
The ladder-type Rydberg EIT is driven by a probe and a coupling fields, coupled $|g\rangle - |e\rangle$ (ground-excited states) and $|e\rangle - |r\rangle$ (excited-Rydberg states) transitions, respectively. 
The Zeeman sublevel spectra of Rydberg EIT in large magnetic fields and laser polarizations have been observed and theoretically simulated by a quantum Monte-Carlo wave-function approach~\cite{Zhang2018}.  
In Ref.~\cite{Wu2017}, an EIT peak height of $10\%$ was studied with an optimum probe intensity and the strongest coupling intensity of 18 W/cm$^2$, where peak height is defined as the difference between the probe transmission at the EIT peak and that without EIT. 
In the present work, we perform the measurements examining the effects of laser polarization, magnetic fields, laser intensities, and optical density (OD). After considering the Zeeman splittings and Doppler shifts, an analytic solution under the perturbation limit of the probe field well simulates the observed spectra and Rydberg EIT properties on the influence of the polarization and magnetic fields.  
We have achieved the maximum Rydberg EIT peak height of $13\%$ after the optimization of the laser intensities and OD. The non-perturbation theoretical approach quantitatively models the EIT properties and predicts well the optimization conditions. More than two-fold improvement of the peak height has been recorded compared to the best results obtained at room temperature. The peak height monotonically improved with the increasing Rabi frequency of the coupling field. 
Limited to our laser source, the strongest coupling power we applied in this work was 27 mW, which corresponds to 5.3 W/cm$^2$ in intensity. 
Using the high-contrast real-time Rydberg EIT spectrum, we will be able to stabilize the laser frequency and traceable detect the environment electromagnetic field as a quantum sensor. 
\medskip

\FigOne

\section{Experimental Setup}
We performed the Rydberg EIT spectral measurements in an Rb vapor cell at the temperature ranging from $27^\circ$C to $65^\circ$C. A probe ($\Omega_p$) and a coupling ($\Omega_c$) fields form a $\Xi$-type EIT configuration. 
The 780-nm probe field (Toptica DL pro) was locked to the transition of the ground state $|5S_{1/2}, F=2\rangle$ to the intermediate state $|5P_{3/2}, F'=3\rangle$ via the sub-Doppler saturation spectroscopy in a reference cell. The 480-nm coupling field (Toptica DL pro HP 480) was swept across the transitions from $|5P_{3/2}, F'=3\rangle$ to $|33D_{3/2}\rangle$ and $|33D_{5/2}\rangle$. 
The numerous magnetic Zeeman sublevels diagram and the schematic of the experimental setup are shown in Figs.~\ref{fig:scheme}(a) and~\ref{fig:scheme}(b). Because the ground and intermediate levels are in linear Zeeman regime under the applied magnetic field in this study, we select $F$ and $m_F=m_J+m_I$ as good quantum numbers, where $m_J$ and $m_I$ are electric angular and nuclear magnetic moment, respectively. The energy splitting of Rydberg states is linearly proportional to the quantum number of $m_J$ and rarely contributed by nuclear magnetic moment $m_I$. The relevant energy splittings of the ground, intermediate, and Rydberg states are 0.70, 0.93, 1.12 (for $|D_{3/2}\rangle$), and 1.68 (for $|D_{5/2}\rangle$) MHz/G between adjacent magnetic sublevels. The arrows in Fig.~\ref{fig:scheme}(a) indicate the most likely transitions driven by the lasers under our experimental conditions. The numbers on the states represent the Clebsch-Gordan coefficients (CGCs) between different Hyperfine and Zeeman states.

The experiments were carried out in a room-temperature or heated vapor cell (Thorlabs GC25075-RB) with the counter-propagating beams, collimated to a full width at $e^{-2}$ maximum of 0.81 mm for both fields. The coupling power was fixed to be 27 mW (expect the measurements applied in Fig.~\ref{fig:ODeffect}(d)) and the probe power was varied from 1.5 $\rm{\mu W}$ to 17 $\rm{mW}$. As illustrated in Fig.~\ref{fig:scheme}(b), two dichroic mirrors (DMs) were used to combine and separate these two color beams. A pair of rectangular Helmholtz coils (11 $\times$ 11 cm$^2$ with 50 windings) generated a homogeneous magnetic field. In the following discussion, we will define the quantum axis in the direction of the magnetic field, which is also the laser beams' propagation direction.
We controlled the laser polarization by two quarter-wave plates ($\lambda/4$). Figure~\ref{fig:scheme}(c) shows the typical Rydberg-state EIT spectrum. The EIT peaks of $|33D_{3/2}\rangle$ and $|33D_{5/2}\rangle$ have the frequency difference of 336.4 MHz~\cite{Mack2011}, which is used to calibrate the coupling field detuning in the whole measurements. All of the spectra were real-time measurements within one second via the PZT scanning of the external cavity diode laser (ECDL), taken directly from the Toptica DLC pro. 

\section{Theoretical model}
We investigate Rydberg EIT in a ladder-type configuration with the following three Hyperfine levels: ground $|g\rangle$, excited $|e\rangle$, and Rydberg $|r\rangle$ states. The atom-field interaction Hamiltonian $H_{int}$ in a rotating frame is represented as ~\cite{Fleischhauer2005}
\begin{equation}
\rm{H_{int}}=\left[
\begin{array}{ccc}
 0 & -\frac{\hbar}{2} \text{$\Omega_p^*$} & 0 \\
 -\frac{\hbar}{2} \text{$\Omega_p$} & \hbar\text{$\Delta_p$} & -\frac{\hbar}{2} \text{$\Omega_c^*$} \\
 0 & -\frac{\hbar}{2} \text{$\Omega_c$} & \hbar\delta \\
\end{array}
\right],
\end{equation}
where $\delta$ is the two-photon detuning defined as $\delta = \omega_p+\omega_c-\omega_{gr}=\Delta_p+\Delta_c$ and 
$\Delta_{p(c)}$ is the detuning of the probe (coupling) laser with frequency $\omega_p$ ($\omega_c$) from the corresponding atomic transition. $\omega_{gr}$ is the transition frequency between the energy levels $|g\rangle$ and $|r\rangle$.
The dynamics of laser-driven atomic systems is governed by the master equation for the atomic density operator
\begin{equation}
\frac{\text{d$\rho $}}{\text{dt}}= \frac{1}{i\hbar} \left[\rm{H_{int}},\rho \right]+\{ \frac{\text{d$\rho$}}{\text{dt}} \}\rm{_{relaxation}},
\end{equation}
where
\begin{equation}
\{ \frac{\text{d$\rho$}}{\text{dt}} \}\rm{_{relaxation}}=\left[
\begin{array}{ccc}
 \Gamma_e\rho_{ee} & -\frac{\Gamma_e}{2}\rho_{ge}  & -(\gamma+\frac{\Gamma_r}{2})\rho_{gr} \\
 -\frac{\Gamma_e}{2}\rho_{eg} & -\Gamma_e \rho_{ee} & -\frac{\Gamma_e+\Gamma_r}{2}\rho_{er} \\
 -(\gamma+\frac{\Gamma_r}{2})\rho_{rg} & -\frac{\Gamma_e+\Gamma_r}{2}\rho_{re} & -\Gamma_r \rho_{rr} \\  
\end{array}
\right].\nonumber
\end{equation}
Here $\Gamma_e$ and $\Gamma_r$ are the spontaneous emissions from the excited and Rydberg states, respectively. $\gamma$ is the coherence dephasing rate between the ground state and Rydberg state. By solving the complete optical Bloch equations (OBEs), we can model the EIT spectrum.
\begin{subequations}
\begin{align}
\partial_{t}\rho_{gg}&=\frac{i}{2}\left(\Omega_{p}^*\rho_{eg}-\Omega_{p}\rho_{eg}^{*}\right)+\Gamma_e\rho_{ee}, \\  
\partial_{t}\rho_{eg}&=\frac{i}{2}\left[\Omega_{p}\left(\rho_{gg}-\rho_{ee}\right)+\Omega_{c}^*\rho_{rg}\right]-\frac{\Gamma_{e}}{2}\rho_{eg}-i\Delta_p\rho_{eg}, \label{eq:4b}\\  
\partial_{t}\rho_{rg}&=\frac{i}{2}\left(\Omega_{c}\rho_{eg}-\Omega_{p}\rho_{er}^{*}\right)-\left(\gamma+\frac{\Gamma_r}{2}+i\delta\right) \rho_{rg}, \\ 
\partial_{t}\rho_{rr}&=\frac{i}{2}\left(\Omega_{c}\rho_{er}-\Omega_{c}^*\rho_{er}^{*}\right)-\Gamma_{r}\rho_{rr},\\  
\partial_{t}\rho_{er}&=\frac{i}{2}\left[\Omega_{p}\rho_{rg}^*+\Omega_{c}^*\left(\rho_{rr}-\rho_{ee}\right)\right]-\left(\frac{\Gamma_e+\Gamma_{r}}{2}-i\Delta_{c}\right)\rho_{er}, \\
1&=\rho_{gg}+\rho_{ee}+\rho_{rr}.  \label{eq4f} 
\end{align}
	\label{eq4}
\end{subequations}
$\rho_{ij}$ is an element of the density-matrix operator of the three-level system and $\rho_{ii}$ represents the population in state $|i\rangle$.
We assume $\Gamma_r \ll \gamma$, $\Gamma_r \ll\Gamma_e$, and further consider $\Omega_p \ll \Omega_c$, the so-called perturbation limit of the probe field. With the assumption, the population stays at the ground state $|g\rangle$, i.e., $\rho_{gg}=1$, $\rho_{ee}=\rho_{rr}=0$. The OBEs in Eq.~(\ref{eq4}) become 
\begin{subequations}
\begin{align}
\partial_{t}\rho_{eg}&=\frac{i}{2}\left(\Omega_{p}+\Omega_{c}^*\rho_{rg}\right)-\frac{\Gamma_e}{2}\rho_{eg}-i\Delta_p\rho_{eg}, \label{eq:5a}\\  
\partial_{t}\rho_{rg}&=\frac{i}{2}\Omega_{c}\rho_{eg}-\left(\gamma+i\delta\right) \rho_{rg}.  
\end{align}
	\label{eq:5}
\end{subequations}
Based on the steady-state solution of the OBEs and Maxwell-Schr$\ddot{\rm{o}}$dinger equation (MSE), we deduce the transmission of the probe field as a function of $\Delta_c$
\begin{equation}
T(\Delta_c)=\exp \left[-\alpha~   \text{Im}\left(\Gamma_e\frac{\text{$\rho_{eg}(\Delta_c)$}}{\text{$\Omega_p$}}\right) \right],
\label{eq:spectrum}
\end{equation}
\rm{where}~~
\begin{equation}
\text{Im}\left(\Gamma_e\frac{\text{$\rho_{eg}(\Delta_c)$}}{\text{$\Omega_p$}}\right)= \frac{\Gamma_e^2 \left(4\gamma ^2+4 \text{$\delta^2$}\right)+2\gamma  \text{$\Omega_c^2$}\Gamma_e}{\left(2\gamma \Gamma_e+\text{$\Omega_c^2$}-4\Delta_p\delta\right)^2+(4\Delta_p \gamma+2 \Gamma_e \text{$\delta$})^2}\equiv\sigma_{eg}(\Delta_c),\\
\end{equation}
\rm{and}~~

\begin{align} \nonumber
\delta &=B\times\left(\Delta_{g}+\Delta_{e}+\Delta_{r}+\Delta_{D}\right)+\Delta_s+\Delta_c,\\ \nonumber
\Delta_{g} &=\frac{\mu_B}{\hbar} g_{F,g}m_{F},\\  \label{eq:shift}
\Delta_{e} &=\frac{\mu_B}{\hbar} g_{F,e}m_{F'},\\  \nonumber
\Delta_{r} &=\frac{\mu_B}{\hbar} g_{J,r}m_J,\\  \nonumber
\Delta_{D} &=(\Delta_{e}-\Delta_{g})\frac{\lambda_p-\lambda_c}{\lambda_c}, \nonumber
\end{align}
under a magnetic field $B$. 
$\sigma_{eg}$ represents the ratio of the absorption cross section to the resonant absorption cross section $\sigma_0$, and hence, $\sigma_{eg}$ is a dimensionless function. $\Delta_s=0$ or 2$\pi\times$336.4 MHz for $|33D_{3/2}\rangle$ or $|33D_{5/2}\rangle$ EIT transition~\cite{Mack2011}. We define the coupling detuning $\Delta_c=0$ at which the $|33D_{3/2}\rangle$ EIT peak occurs. Here $\delta$ is a function of the applied magnetic field due to the Zeeman effect and Doppler shift. $\Delta_g$, $\Delta_e$, and $\Delta_r$ are the Zeeman shifts of ground, excited, and Rydberg energy levels, respectively. $g_{F,g}=1/2$, $g_{F,e}=2/3$, and $g_{J,r}=0.8$ (for $|D_{3/2}\rangle$ state) or $g_{J,r}=1.2$ (for $|D_{5/2}\rangle$ state) are the corresponding Land\'e g-factors. An additional Doppler shift $\Delta_D$, which considers the selection of the velocity group, plays a role in the energy splitting under the magnetic field~\cite{Zhang2018}. 
$\alpha$ is the optical density and $\Gamma_e$ is the linewidth of the excited state $|5P_{3/2}\rangle$, which in principle is $2\pi\times 6$ MHz but will be adjusted in the simulation with $\Omega_p$-varied measurements in Sec.~\ref{sec:optimization}. 

For each EIT spectral measurement, we locked the frequency of the probe laser and swept the coupling laser frequency. A typical spectrum is shown in Fig.~\ref{fig:scheme}(c) with a small peak of $|33D_{3/2}\rangle$ and a large peak of $|33D_{5/2}\rangle$ Rydberg EITs. 
The coupling strength between two Hyperfine levels is calculated by the dipole matrix element, which is a product of CGC and a reduced matrix element, using the Winger-Eckart theorem. The ratio of the reduced matrix element of $|33D_{3/2}\rangle$ and $|33D_{5/2}\rangle$ is around 1/3~\cite{Steck}. 
We set $\Delta_p=0$, $B=0$ and assume the populations are equally distributed in states $|1\rangle$ and $|6\rangle$. The determined $\alpha$ is 1.65 based on the baseline transmission and $\{\Omega_c, \gamma\}=2\pi\times\{7.8, 5.4\}$~MHz according to the EIT spectral profile.
The analytic solution can describe the Doppler-broadened EIT spectrum with the effective parameters so far, and it will be applied to investigate the influence of the laser polarization and magnetic field. 
Note that to enhance the Rydberg EIT peak height, the utilized probe intensity was around $0.04~\rm{W/cm^2}$ (or $\Omega_p = 2\pi\times30$~MHz in the theoretical model of Sec.~\ref{sec:optimization}) which is not sufficiently weak. The simulation under a non-perturbation calculation still works with the set of effective $\Omega_c$, $\gamma$, and $\Gamma_e$ parameters. 

\section{Rydberg EIT spectral measurements}
We then present Rydberg-state EIT spectral measurements on the effects of the laser polarization, magnetic field, laser intensity, and optical density of the medium. After optimization, a peak height of $13\%$ has been achieved. Our theoretical calculation supports the experimental observations and predicts that the peak height is not limited to the present result. The details will be discussed in Sec.~\ref{sec:optimization}.

\FigTwo

\subsection{Evolution of Rydberg EIT along with laser polarization}
Polarization configurations of the probe and coupling fields must be well adjusted according to Rydberg $S$- or $D$-orbital states. The expression of $T=\exp \left(\frac{-2\alpha \gamma\Gamma_e}{2\gamma\Gamma_e+\Omega_c^2} \right)$ derived from Eq.~(\ref{eq:spectrum}) indicates that a stronger $\Omega_c$ corresponds to a higher EIT peak. The deviation of the effective Rabi frequency requires consideration of all possible transitions between the Hyperfine and Zeeman states in various laser polarization configurations. In our measurements, the probe field was maintained in the $\sigma^+$-polarization while that of the coupling field was adjusted by a $\lambda/4$ wave plate. The evolution of the multiple EIT structure along with the coupling field polarization (data were taken every 10 degrees of the wave plate) under a magnetic field of 10.9 Gauss is shown in Fig.~\ref{fig:2}. We assume all populations are accumulated at state $|1\rangle$ as a result of the optical pumping effect caused by the $\sigma^+$ probe field. According to the selection rules, we only consider three possible transitions involving states $|1\rangle$ and $|2\rangle$, shown on the right-hand side of Fig.~\ref{fig:scheme}(a). In addition, the relevant CGCs of the coupling field, the Zeeman splitting, the Doppler shift, and the ratio of $\sigma^+$ to $\sigma^-$ of the coupling field are considered. 
The total ratio of the absorption cross section, $\sigma_{eg}$, is the product of individual one, i.e., 
\begin{equation}
\sigma_{eg}=\prod _{n=3,4,5} \text{$\sigma_{2n}$},\\
\end{equation}    
where $\sigma_{2n}$ represents the EIT absorption cross-section involving states $|2\rangle$ and $|n\rangle$. From the top to bottom in Fig.~\ref{fig:2}(a), the black dashed lines show the measurements with the $\sigma^+$, linear, and $\sigma^-$ coupling field polarizations, and the red solid lines indicate the corresponding fits. 

We further extract the three EIT peak heights in each measurement and plot them in \ref{fig:2}(b). Black squares, blue circles, and green triangles show the peak heights involving the transitions of $|1\rangle\rightarrow|2\rangle\rightarrow|3\rangle$,  $|1\rangle\rightarrow|2\rangle\rightarrow|4\rangle$, and $|1\rangle\rightarrow|2\rangle\rightarrow|5\rangle$, respectively. 
The black solid line is the simulation result under a parameter set of $\alpha=1.55$, $\{\Omega_c$ (for CGC=1), $\gamma\}=2\pi\times\{5.2,8\}$MHz, which fits the heights of the main (right) peak of the $|33D_{5/2}\rangle$ EIT. 
For the $|33D_{3/2}\rangle$ and the left peak of $|33D_{5/2}\rangle$ EITs, the corresponding coupling Rabi frequencies (after calculating the reduced matrix element and the relevant CGC) need to be enlarged 1.5 fold. 
The transmission equation in Eq.~(\ref{eq:spectrum}) does not consider the contribution from different velocity groups; however, the Doppler effect induces a broadening of the EIT linewidth and a reduction of the peak height. Therefore, the 1.5-fold calibration of $\Omega_c$ for the secondary EITs is reasonable. We apply the same calibration in the simulation of Fig.~\ref{fig:3}, which shows the EIT spectra in different magnetic fields. 
Therefore, for the given $nD$ Rydberg state, we selected the  $\sigma^+$-$\sigma^+$ polarization configuration to get the best EIT contrast. 

\subsection{Evolution of Rydberg EIT along with the magnetic field}
\FigThree
The analytic function shown in Eq.~(\ref{eq:spectrum}) can interpret the rich structure of multiple EITs in magnetic fields, including peak positions, height, and linewidth of the EITs. The magnetic field was generated by a pair of rectangular Helmholtz coils in the y axis, which was parallel to the direction of the laser propagation.
We then varied the magnetic field from 4 to 22 Gauss at the temperature of $27^\circ$C to achieve a high resolution of the frequency splitting. The probe and coupling fields were linearly polarized along the z-axis, which is perpendicular to the quantum axis defined in the magnetic field direction. Therefore, these two lasers both have equal $\sigma^+$ and $\sigma^-$ polarized components.
The measurements (black dashed lines) and simulations (red solid lines) of Rydberg EIT spectra with magnetic fields of 7.3, 15, and 22 Gauss are shown in the panel of Fig.~\ref{fig:3}(a). 
Here, we assume the populations are equally distributed in states $|1\rangle$ and $|6\rangle$. Due to the larger CGCs, the right and left main peaks correspond to transitions of $|1\rangle \rightarrow |2\rangle \rightarrow |3\rangle$ and $|6\rangle \rightarrow |7\rangle \rightarrow |8\rangle$, respectively. Here, we consider six possible transitions, three of them involve states $|1\rangle$ and $|2\rangle$, while the other three involve states $|6\rangle$ and $|7\rangle$. The relevant CGCs of the coupling field, the Zeeman splitting, and the Doppler shift are taken into account. Again, $\sigma_{eg}$ is the product of individual one, 
\begin{equation}
\sigma_{eg}=\prod _{n=3,4,5} \text{$\sigma_{2n}$}\prod _{n=8,9,10} \text{$\sigma_{7n}$},\\
\end{equation}    
where $\sigma_{2n}$ and $\sigma_{7n}$ represent the EIT absorption cross-section involving states $|2\rangle$-$|n\rangle$ and $|7\rangle$-$|n\rangle$, respectively. To well fit the data, the set parameters are $\alpha=0.36\pm 0.02$ according to the baseline transmission and $\{\Omega_c$ (for CGC=1), $\gamma\}=2\pi\times\{2.1, 6\}$MHz based on the EIT profile. Note that the above-mentioned 1.5-fold calibration of $\Omega_c$ for the secondary EITs is also performed. In the spectrum, there are six resonant EIT peaks, each located at a different detuning of the coupling field. Take $|33D_{5/2}\rangle$ Rydberg EIT as an example. The four peaks from left to right are the transitions of $|6\rangle \rightarrow |7\rangle \rightarrow |8\rangle$, $|6\rangle \rightarrow |7\rangle \rightarrow |9\rangle$, $|1\rangle \rightarrow |2\rangle \rightarrow |4\rangle$, and $|1\rangle \rightarrow |2\rangle \rightarrow |3\rangle$. Therefore, Eq.~(\ref{eq:spectrum}) with the detunings in Eq.~(\ref{eq:shift}) can qualitatively and quantitatively describe the Rydberg EIT spectra, peak height, and the relevant physics properties such as the Zeeman splitting. The EIT linewidth was around 10 MHz, which enables the analysis of the rich Rydberg EIT structure with an increasing magnetic field. As the applied magnetic field is sufficiently large, e.g. 6 Gauss, the outermost peaks can be resolved fully. The 22-Gauss asymmetry spectrum was caused by the satellite peaks relative to the main peaks, suggesting that we need to consider more possible transitions not only involving state $|1\rangle$ and state $|6\rangle$~\cite{Zhang2018}. 

The frequency shifts of the EIT peaks are further extracted as functions of the magnetic field. The black squares and green circles in Fig.~\ref{fig:3}(b) represent the frequency splittings of two outermost and two innermost peaks of the EITs, respectively. The linear fits give the frequency splittings of 7.5 and 0.77 MHz/G. In the calculation as discussed in Eq.~(\ref{eq:shift}), the shift of the two main EIT peaks cased by Zeeman effect is $2\times(\Delta_r|_{m_J=5/2}-\Delta_g|_{m_F=2})=2\times2.8$ MHz/G without considering the atom velocity. The Doppler shift $\Delta_D$ contributes an additional splitting, $2\times(\Delta_e|_{m_{F'}=3}-\Delta_g|_{m_F=2})\frac{\lambda_p-\lambda_c}{\lambda_c}=2\times 0.87$ MHz/G.  
Accordingly, the total frequency splitting for these two main peaks is 7.3 MHz/G, and similarly, that of the innermost EITs is 0.62 MHz/G. 
The minor discrepancy would be the inhomogeneous magnetic field generated by the Helmholtz coils. The field at the edge of the cell is $30\%$ stronger than that in the center. Thus, this research provides an important prerequisite for high-contrast spectroscopic EIT in sites with magnetic backgrounds. 
We further show the EIT peak height and its relevant EIT linewidth under $\sigma^+$-$\sigma^+$ polarization configuration and temperature of $51^{\circ}$C in Fig.~\ref{fig:3}(c). The maximum peak height, as well as the narrowest linewidth, occurred when the magnetic field was around zero. We discovered that the non-uniform magnetic field also causes a broader and lower EIT spectrum. 

\subsection{Optimization of Rydberg EIT peak height} \label{sec:optimization}
The polarization configurations of the laser fields were adjusted to be $\sigma^+$-$\sigma^+$ and the magnetic field was tuned to be around zero Gauss to achieve the best EIT contrast, which has the maximum peak height and narrowest linewidth. We then systematically optimized the Rydberg EIT peak height by varying the intensity of the probe field ($I_p$) and optical density ($\alpha$) by heating the vapor cell. The baseline transmission $\rm{T_{B}}$ (the probe transmission without applying the coupling field) and the EIT peak height (defined as the transmission difference between the peak and baseline) are shown in Figs.~\ref{fig:ODeffect}(a) and ~\ref{fig:ODeffect}(b). As discussed in Ref.~\cite{Wu2017}, the power broadening effect slightly raises the baseline transmission and the coherent population transformation from $|g\rangle$ to $|r\rangle$ also monotonously improves the peak transmission of the EIT when increasing $I_p$. For any given temperature, the peak height has a universal behavior that it reaches the maximum value with the optimum $I_p$. The optimum $I_p$ becomes stronger with a higher vapor temperature, as shown in Fig.~\ref{fig:ODeffect}(b). We achieved the maximum EIT peak height of 13$\%$ with the vapor temperature of $51^{\circ}$C and $I_p$ of 0.044 W/cm$^2$.  

\FigFour

By applying the theoretical model, we are able to predict the universal behavior. First of all, we discuss the baseline transmission increases with $I_p$ which is mainly caused by the power broadening effect, as expressed by $\rm{T_B}= \exp \left(-\alpha\frac{  \text{$\Gamma_e^2$}}{\text{$\Gamma_e^2$}+2 \text{$\Omega_p^2$}}\right)$. We extract the optical density $\alpha$ and the conversion ratio of laser intensity to $\Omega_p/\Gamma_e$ in a series of measurements with different $I_p$. The determined $\alpha$ varied from 0.42 ($27^{\circ}$C) to 5.0 ($65^{\circ}$C) and $\Omega_{p,\rm{1W/cm^2}}$ (for $I_p=1 \rm{W/cm^2}$) also varied from 3.9 $\Gamma_e$ to 1.8 $\Gamma_e$ from low to high temperature. Below shows the possible explanations for the varied intensity conversion ratios. The probe field can optically pump and redistribute the population among five Zeeman states of $|5S_{1/2}, F=2\rangle$. The optical pumping would be inefficient in a dense medium condition due to the radiation trapping effect~\cite{Rosenberry2007,Thomas2017} and the short transit time that atoms move in and out of the interaction regime. 
The Rabi frequency of the probe field is derived as 
$\Omega_p= \langle a_p\rangle \Gamma~\sqrt[]{\frac{I_{p}}{2 I_{sat}}}$, where $\langle a_p\rangle=1$ (or $\sqrt{7/15}$) if all populations are pumped to single Zeeman state $|1\rangle$ (or all populations are equally distributed between five Zeeman states of $|5S_{1/2}, F=2\rangle$). In the expression of $\rm{T_B}$, these effects are not taken into account. 
From the EIT spectra, $\Omega_c$ and $\gamma$ can be resolved by the best fits of the profile (including the peak height and linewidth of the EIT) by applying Eq.~(\ref{eq:spectrum}). However, for a given temperature, the determined effective $\Omega_c$ and $\gamma$ do not maintain constant values as varied $I_p$. Instead, both values monotonously increase with $I_p$. For these measurements, the simplified equation, which assumes $\Omega_p\ll \Omega_c$ and $\rho_{gg}=1$, does not work properly. 

In the following simulation, we applied a non-perturbation numerical calculation by solving all of the Eqs.~(\ref{eq4}) and ~(\ref{eq:spectrum}). In the whole simulation shown in Fig.~\ref{fig:ODeffect}(b), we set $\Gamma_r/2\pi = 5$ kHz and fixed $\Omega_c/2\pi = 0.38\pm 0.03 $ MHz and $\gamma/2\pi = 0.75$ MHz. The optical density and the conversion ratio of $I_p$ to $\Omega_p/\Gamma_e$ are first derived by the best fits to the baseline transmission in Fig.~\ref{fig:ODeffect}(a). By comparing the theoretical EIT linewidths to the spectral measurements, $\Gamma_e/2\pi$ is determined to be $60$ MHz, which includes the selection of velocity groups participating in the interaction. The optimum $I_p$ and peak heights are accurately predicted in the simulation for each temperature (optical density). 
We further extract the maximum EIT peak height and the relevant linewidth with the determined optical density from the different vapor temperatures. The data were interpolated from the measurements in Fig.~\ref{fig:ODeffect}(c). The solid lines are calculated using the above-mentioned parameters and a fixed $\Omega_c/2\pi = 0.38$ MHz. 
The Doppler-free non-perturbation model well describes the measurements, including the baseline transmission, peak height, and linewidth of EIT. It can also phenomenologically predict the maximum EIT peak height based on the OD and the probe field intensity. Therefore, our measurements and the simulation provide a better way to study and optimize the high contrast Rydberg-EIT spectrum, which will advance Rydberg-atom-relevant research. 
Moreover, the height can be further enhanced with a strong coupling field, so that it is not limited to the present study. We prove the principle by reducing the coupling laser power. As shown in Fig.~\ref{fig:ODeffect}(d), the peak height monotonously increases with $I_c$ and the applied maximum coupling intensity is far from the saturation of the EIT peak transmission. 
The measurements are consistent with the theoretical prediction (solid line).
As a result of the increased $I_c$, the linewidth only varied only by 10$\%$. Therefore, the comprehensive feature of EIT including the peak height and linewidth leads to a better sensitivity for frequency locking of the upper transition and for traceable detection of the environment electromagnetic field as a quantum sensor.

\section{Conclusion}
We studied the Rydberg EIT spectra in different polarizations of the light fields and external magnetic fields. The spectrum can be described by an  analytic solution using effective parameters that include the Doppler effect. The function can also explore the rich structure of multiple EITs in magnetic fields, including the peak resonant frequency, height, and linewidth of the EIT. From the detection of EIT structure or frequency splitting, this study provides an important prerequisite for high-contrast spectroscopic EIT in sites with magnetic backgrounds. Additionally, we have demonstrated the enhancement of the EIT peak height through the optimization of laser intensity and optical density. After optimization, the maximum peak height of 13$\%$ was achieved. By utilizing a non-perturbation numerical calculation, we can qualitatively and quantitatively model the EIT behavior, including the EIT peak height and linewidth, as well as the optimum probe intensity at various optical densities. With the use of a stronger coupling field, the EIT contrast can be further improved. We expect our study of Rydberg EIT enhancement to have implications for future applications.
\begin{backmatter}

\bmsection{Acknowledgments}
This work was supported by the Ministry of Science and Technology of Taiwan under Grant Nos. 109-2112-M-110-008-MY3, 110-2123-M-006-001, and financially supported by the Center for Quantum Technology from the Featured Areas Research Center Program within the framework of the Higher Education Sprout Project by the Ministry of Education (MOE) in Taiwan. The authors thank Professor Ite A. Yu for valuable comments on the paper.
\bmsection{Disclosures} The authors declare no conflicts of interest

\end{backmatter}

\bibliography{RydbergEIT}


\end{document}